# From 'Individual Scientist' to 'Integrated Scientist': The Evolution of Scientific Organizational panels and Their Impact on the Scientific System

Zekai Zhang

zzk@zjut.edu.cn

Abstract: This article aims to propose and elucidate the analytical concepts of "individual scientist" and "integrated scientist" to depict the fundamental transformation in the modes of scientific research actors throughout the history of science. The "individual scientist" represents an early modern scientific research panel characterized by independence, egalitarian collaboration, and personal recognition, while the "integrated scientist" emerged in the context of "big science," marked by hierarchical teams, division of labor, collaboration, and the concentration of recognition on team leaders. Through historical review and case analysis, this article explores the underlying drivers of this transformation and focuses on its challenges and reconstructions concerning the name-based scientific reward system, aiming to provide a reflective perspective for contemporary scientific governance and research evaluation.

## 1. Introduction: The "Two Ideal Types" of Scientists

In public imagination, science is often portrayed as the flash of insight of a solitary genius, from Newton's contemplations at Woolsthorpe Manor to Einstein's "miracle year" at the Bern patent office. This image is what this article defines as the "Individual Scientist." Its characteristics include research activities that heavily depend on personal intellect and resources, collaboration based on equal exchange of shared interests, and clear attribution of scientific discoveries to the individual.

However, since the mid-20th century, another image of scientists has increasingly emerged: they are leaders or members of large teams, conducting research that relies on expensive public equipment and continuous massive funding, with results co-signed by dozens or even hundreds of people. This is the "Integrated Scientist"—someone incorporated into a formal, hierarchical research system. This article argues that the shift from "Individual Scientist" to "Integrated Scientist" is not merely an expansion of scientific scale but a profound revolution in the social organization of science, fundamentally reshaping the logic of scientific production and the distribution of honors.

## 2. The Historical View of the 'Individual Scientist' Paradigm

The 'individual scientist' paradigm corresponds to the era of 'small science.' before 20th century. Typical cases including Galileo observing celestial bodies with a homemade telescope; Mendel conducting pea hybridization experiments in the monastery garden; Darwin compiling decades of data to write "*The Origin of Species*". Their research did not require large teams or national-level budgets, and basically on their own interest, and show the social structural characteristics below:

1. Independence of research: Individual scientists could control the entire process from raising questions, designing experiments, to constructing theories.

2. Equality in collaboration: Communication and letters were the main forms of collaboration. For example, the famous 'saddler and scientist' partnership between Boyle and Hooke, though involving employment, was more like academic partnership.

3. Individual fame: The core of the scientific reward system was the competition for priority, with clear authorship and unambiguous attribution of honor. The dispute over calculus between Newton and Leibniz is an extreme example of this individual sense of honor.

Its core is built on the 'scientific ethos' described by Merton, particularly disinterestedness (researching for the sake of science itself) and communalism (results belong to the entire scientific community).

## 3. The Rise of the 'Integrated Scientist': Motivations and Characteristics

The emergence of the 'integrated scientist' is an inevitable result driven by multiple factors with the society evolution:

1.Technological push: Experimental science is becoming increasingly complex, requiring interdisciplinary knowledge and cutting-edge equipment (such as particle colliders and gene sequencers), far beyond the capacity and financial means of any individual.

2. Institutional demands of 'big science': After World War II, states became the main funders of scientific research. To meet national strategic needs (such as the Manhattan Project and the Moon landing program), scientific activities needed to be organized and managed like industrial projects, emphasizing efficiency, division of labor, and goal orientation.

3. Advancement of scientific knowledge: The complexity of scientific problems requires high specialization, and no individual can master all the necessary knowledge, making it essential to form teams with each member performing specific roles.

The 'integrated scientist' exhibits characteristics that are markedly different from those of their predecessors:

1. Collective nature of research: Scientific research becomes a 'project' completed by teams. Individuals (especially junior researchers) often handle only a single link in the entire knowledge production chain.

2. Hierarchical status: Clear power hierarchies exist within laboratories or project teams. The PI or chief scientist is at the top, controlling direction, resources, and the distribution of honors; beneath them are young teachers, postdocs, graduate students, and technicians in subordinate positions.

3. Centralization and ambiguity of recognition: External recognition (awards,

media attention, academic influence) mainly goes to the team leader. The 'corresponding author' system somewhat solidifies this pattern, as the PI is seen as the 'owner' of the results even if they did not personally perform the experiments. while in multi-author papers, the specific roles of contributors other than the first and corresponding authors are hard to identify, resulting in a mismatch between contributions and rewards.

## 4. The Profound Impact of Paradigm Shifts: Examining the Science Reward System

The shift from the "individual scientist" to the "integrated scientist" has profoundly impacted the core of science—the reward system.

1. Alienation of Authorship: Authorship has evolved from a pure symbol of honor into a form of "currency" that affects one's career. Under the pressure of "publish or perish," distorted phenomena have emerged, such as the "honorary author" (senior individuals with minimal contributions) and the "ghost author" (junior researchers who made key contributions but were overlooked).

2. Challenges in the Evaluation System: Traditional evaluation metrics based on the number of publications and citation counts have become inadequate in the era of "integrated science." They struggle to accurately measure the contributions of individual team members, thereby exacerbating the "Matthew effect," strengthening the already powerful while systematically undervaluing young scholars and behind-the-scenes contributors

3. Training and Loss of Scientific Talent: Hierarchical structures may cause junior researchers to feel alienated, seeing themselves as "research laborers" rather than explorers of knowledge. This career experience can dampen scientific enthusiasm and lead to talent attrition.

4. Potential Risks to the Innovation Ecosystem: When honors and resources are overly concentrated among a few "star scientists," the survival space for independent, high-risk, and unconventional "small science" research may be squeezed, which in the long run is detrimental to the diversity of scientific thought.

## 5. Discussion: Reshaping the Spirit of Science in the Era of 'Integration'

The 'integrated scientist' panel is an inevitable development in science, bringing unprecedented research capabilities, but its inherent tensions should not be overlooked. We are not trying to return to the romantic era of the 'individual scientist'; rather, we need to innovate institutionally under the new paradigm to maintain a healthy ecosystem for science.

1. Promote contributor credit models: Replace simple author lists with more refined systems such as CRediT (Contributor Roles Taxonomy), clearly indicating each individual's specific contributions in conceptualization, methodology, experiments, data analysis, manuscript writing, and other aspects.

2. Reform research evaluation mechanisms: When assessing individuals, emphasis should be placed on their specific contributions rather than simply on author position or number of publications. Encourage open science practices and recognize non-traditional outputs such as data sharing and code development.

3. Strengthen academic ethics and lab culture: Advocate for PIs to take greater responsibility in mentorship and establish a culture within teams that is fair, transparent, and recognizes the value of all members.

## 6. Conclusion

The transformation from the 'individual scientist' to the 'integrated scientist' outlines the modernization process of science as a social institution. While this shift enhances scientific productivity, it also poses significant challenges to its inherent spirit and reward system. Understanding these two 'ideal types' helps us more clearly examine the operational logic of contemporary science and actively construct a future scientific landscape that can harness the power of integration while safeguarding individual dignity and creativity.

# References

1. Merton, R. K. (1973). The Sociology of Science: Theoretical and Empirical Investigations. University of Chicago Press.

2. Price, D. J. de S. (1963). Little Science, Big Science. Columbia University Press.

3. Ziman, J. (2000). Real Science: What it is, and what it means. Cambridge University Press.

4. Stephan, P. (2012). How Economics Shapes Science. Harvard University Press.